\documentclass[conference]{IEEEtran}
\IEEEoverridecommandlockouts
\usepackage{cite}
\usepackage{amsmath,amssymb,amsfonts}
\usepackage{algorithmic}
\usepackage{textcomp}
\def\BibTeX{{\rm B\kern-.05em{\sc i\kern-.025em b}\kern-.08em
    T\kern-.1667em\lower.7ex\hbox{E}\kern-.125emX}}

\usepackage{xspace}
\usepackage{soul}
\usepackage{mathtools}
\usepackage{makecell}
\usepackage{cite}
\usepackage{nicefrac}
\usepackage{float}
\usepackage{wrapfig}
\usepackage{amssymb}
\usepackage{amsmath}
\usepackage[dvipsnames]{xcolor}
\usepackage{graphicx}
\usepackage{subcaption}
\usepackage[ruled,vlined]{algorithm2e}

\newcommand{\ra}{\rightarrow}

\newtheorem{asm}{Assumption}
\newtheorem{prp}{Proposition}
\newtheorem{thm}{Theorem}
\newtheorem{rem}{Remark}

\newcommand{\ie}{\unskip, i.\,e.,\xspace}
\newcommand{\eg}{\unskip, e.\,g.,\xspace}
\newcommand{\pd}{p.\,d.\xspace}

\newcommand{\wrt}{w.\,r.\,t.\xspace}

\newcommand{\N}{\ensuremath{\mathbb{N}}}

\newcommand{\R}{\ensuremath{\mathbb{R}}}

\newcommand{\U}{\ensuremath{\mathbb{U}}}

\newcommand{\eps}{\ensuremath{\varepsilon}}

\DeclareMathOperator*{\arginf}{arg\,inf}

\makeatletter
\newcommand{\subalign}[1]{%
	\vcenter{%
		\Let@ \restore@math@cr \default@tag
		\baselineskip\fontdimen10 \scriptfont\tw@
		\advance\baselineskip\fontdimen12 \scriptfont\tw@
		\lineskip\thr@@\fontdimen8 \scriptfont\thr@@
		\lineskiplimit\lineskip
		\ialign{\hfil$\m@th\scriptstyle##$&$\m@th\scriptstyle{}##$\crcr
			#1\crcr
		}%
	}
}

\usepackage{ marvosym }
\begin{document}

\title{On performance bound estimation in NMPC with time-varying terminal cost}

\author{\IEEEauthorblockN{Lukas Beckenbach}
\IEEEauthorblockA{\textit{Automatic Control and System Dynamics Laboratory} \\
\textit{Technische Universität Chemnitz}\\
09107 Chemnitz, Germany \\
lukas.beckenbach@etit.tu-chemnitz.de}
\and
\IEEEauthorblockN{Stefan~Streif}
\IEEEauthorblockA{\textit{Automatic Control and System Dynamics Laboratory} \\
\textit{Technische Universität Chemnitz}\\
09107 Chemnitz, Germanyy \\
stefan.streif@etit.tu-chemnitz.de}
}

\maketitle

\begin{abstract}
Model predictive control (MPC) schemes are commonly designed with fixed \ie time-invariant, horizon length and cost functions.
If no stabilizing terminal ingredients are used, stability can be guaranteed via a sufficiently long horizon. 
A suboptimality index can be derived that gives bounds on the performance of the MPC law over an infinite-horizon (IH).
While for time-invariant schemes such index can be computed offline, less attention has been paid to time-varying strategies with adapting cost function which can be found, e.\,g., in learning-based optimal control. 
This work addresses the performance bounds of nonlinear MPC with stabilizing horizon and time-varying terminal cost. 
A scheme is proposed that uses the decay of the optimal finite-horizon cost and convolutes a history stack to predict the bounds on the IH performance.
Based on online information on the decay rate, the performance bound estimate is improved while the terminal cost is adapted using methods from adaptive dynamic programming.
The adaptation of the terminal cost leads to performance improvement over a time-invariant scheme with the same horizon length.
The approach is demonstrated in a case study. 
\end{abstract}


\section{Introduction}

Optimal control schemes such as model predictive control (MPC) are used when control performance is a design criterion. 
In MPC, a cost function is optimized at each time step over a finite-horizon \cite{Gruene2017}.
Asymptotic stability of a set point is commonly guaranteed using either terminal ingredients (such as a terminal constraint set, e.\,g., \cite{Chen1998}) or a sufficiently long prediction horizon (e.\,g.~\cite{Limon2006}). 

While finite-horizon (FH) formulations are tractable, applying their controller obviously leads to performance loss on a long horizon compared to the (generally intractable) infinite-horizon (IH) formulation.
Under the so-called asymptotic controllability assumption \cite{Gruene2009,Reble2011} the prediction horizon can be associated with a suboptimality index, which relates the performance of the FH optimal controller and the IH performance.
Using this framework, several performance studies have been carried out in different settings.
In \cite{Lu2020}, an event-triggered scheme is proposed which preserves initial suboptimality under a sufficiently long (stabilizing) prediction horizon. 
The event-triggering entails reduction of the computational effort as more than one element of the optimal FH open-loop controls can be applied.
A performance estimation approach is proposed in \cite{Gruene2009a}, where a priori and a posterio bounds on the performance of \emph{time-invariant} MPC can be estimated via numerical information along the closed-loop trajectory. 

Model uncertainties can decrease controller performance.
Approaches which directly take into account model uncertainties and which guarantee, e.~g., constraint satisfaction in a probabilistic or worst-case setting, are stochastic or robust MPC (see \eg \cite{Mesbah2014,Magni2003}). 
Although the control performance can be increased by improving on the model accuracy (see, e.~g.~\cite{Piga2019}), there is a trade-off between learning a good model and learning the optimal cost \cite{Recht2019}.
Another and direct approach to reduce the influcence of uncertainties on control performance is to keep the prediction horizon short. 
In this spirit, \cite{Pannek2011} analysed the behaviour of MPC with shorter horizon and investigated the suboptimality if more than one open-loop control elements are applied to the system. 
In, e.\,g.~\cite{Tuna2006} a condition for a stabilizing horizon length is shown assuming a relaxed control Lyapunov function type terminal cost.

Stage and terminal cost designs and adaptation schemes have been proposed to improve controller performance in a FH setting.
The work \cite{Hu2002} aims at IH optimality and uses MPC with terminal costs that involve a local Lyapunov function as well as a time-invariant weight on the terminal predicted state away from the vicinity of the origin. 
Other works are e.\,g.~\cite{Coron2020} or \cite{Beckenbach2020a} in which the stage cost is modified using tools such as homogeneous approximations or reinforcement learning techniques, respectively.
In \cite{Alamir2018}, the stage cost has been designed as a monotonically increasing penality over the prediction horizon. 
Cost tuning has also been proposed in \cite{DiCairano2010} such that the controller matches a specified linear control.
The terminal cost is adapted in \cite{Rosolia2018} in an iterative learning fashion based on collected history data. 
Furthermore, \cite{Koehler2021} shows the benefit of using a terminal cost function constructed as a FH tail given a local exponentially stabilizing controller to reduce the required prediction horizon for stabilization and offers a performance estimate.

In MPC schemes, in which the cost is adapted or time-varying, the IH performance is not easily computable a priori unless further assumptions are made.
Such assumptions include, for example, establishing uniform bounds on the optimal FH cost over all time steps (see \cite[Assumption 6.29]{Gruene2017}).
This may, however, potentially yield conservative results as the step-wise performance is not specifically accounted for.
The so called dynamic regret analysis (see, e.g.~\cite{Zhang2021}  and references therein) offers an online tuning design for weights of the stage cost aiming at cost reduction by exploiting the online obtained information. The approach is based on a decomposed accumulated (hypothetical) error between the generated control action and the IH optimal controller.

This work is devoted to an unconstrained MPC scheme with adapting thus time-varying terminal cost.
An estimate of the IH performance is provided based on initial information about the suboptimality and by relating the FH optimal cost decay to a convolution of past decay rates. 
Aiming to approximate the IH tail, the terminal cost is modified but not limited to using methods of adaptive dynamic programming (ADP).
The terminal cost depends -- albeit not explicitly -- on time, as is also often the case in online learning-based schemes. 
It is shown that if the time-varying FH cost fulfills a specific convoluted decay, the performance can be bounded a priori.
As the collected decay depends on the trajectory taken, this a priori statement is conservative. 
Therefore, an online performance bound correction is proposed that improves the a priori bound estimate.
The key aspect of the estimation lies in the possibility to project the accumulated point-wise cost up to the current time step onto the IH performance via convolution.
Although the online computational load is slightly increased, the proposed scheme is at least as good as the time-invariant scheme with the same prediction horizon.

\section{Problem Formulation and Preliminaries}

Consider the discrete-time system dynamics
\begin{align} \label{eq:sys}
	x^+ = f(x,u), 
\end{align} 
where $x \in \R^n$ and $u \in \R^m$ represent the systems state and input, respectively, and $f:\R^n \times \R^m \ra \R^n$ is a continuous transition map. 
Without loss of generality it is assumed that $f(0,0) = 0$. 

Given an initial point $x^o \in \R^n$ and a control sequence $\{u(k;x^o)\}_{k \in \N_0}$ associated to $x^o$, let $\{x_u(k;x^o)\}_{k \in \N_0}$ denote the resulting states of \eqref{eq:sys}, satisfying $x_u(0;x^o) =x^o$ and $x_u(k+1;x^o) = f(x_u(k;x^o),u(k;x^o))$. 

\subsection{N-horizon control problem}

A finite-horizon (FH) control $u(\cdot;x) \in \R^{m \times N}$, with prediction horizon $N \in \N$, is said to be optimal if it minimizes the FH cost function
\begin{align} \label{eq:FH-cost}
	J_{N}(x,u(\cdot;x)) = \sum_{k=0}^{N-1} l(x_u(k;x),u(k;x))
\end{align}
associated with the \pd\footnote{A function $h:\R^n \ra \R_{\geq 0}$ is called positive-definite (\pd) if $h(x)>0$ for all $x \neq 0$ and $h(0)=0$.} tracking cost, or utility, $l: \R^n \times \R^m \ra \R_{\geq 0}$ satisfying
\begin{asm} \label{asm:track-utility}
	There exists an $\alpha_l \in \mathcal{K}_{\infty}\footnote{A function $h:\R_{\geq 0} \ra \R_{\geq 0}$ is of class $\mathcal{K}_{\infty}$ if $h(0)=0$, $h$ is strictly monotonically increasing and $\lim_{r \ra \infty} h(r) = \infty$.}$ such that for all $x \in \R^n$, $\alpha_l(|x|) \leq \inf_{u \in \R^m} l(x,u)$, where $|\cdot|$ denotes the 2-norm.
\end{asm}
Let $
V_N(x) = \inf_{u(\cdot;x) \in \R^{m \times N}} \; J_{N}(x,u(\cdot;x))
$
denote the optimal FH cost, associated with the optimal control $u^\ast(\cdot;x)$. 
For proper $V_{\infty}$, there exists a \pd function $\alpha_V:\R_{\geq 0} \ra \R_{\geq 0}$ satisfying $V_{\infty}(x) \leq \alpha_V(|x|)$.   
The implicit MPC law, applied to the system, is defined as $\kappa_N(x) := u^\ast(0;x)$. 

It holds that $V_N(\cdot) \leq V_M(\cdot)$, for any $N,M \in \N$ satisfying $N \leq M$ \cite{Rawlings2017}. 

It is known from the theory of relaxed dynamic programming (RDP; \cite{Lincoln2006,Gruene2008}) that for any $\Omega \subset \R^n$ there exists $\underline{N} \in \N$ such that for all $N \geq  \underline{N}$ and all $x \in \Omega$,
\begin{align} \label{eq:RDP-FH-plain}
	V_N(x) + \alpha l(x,u^\ast(0;x)) \leq V_N(f(x,u^\ast(0;x)))
\end{align}
for some $\alpha \in (0,1]$. 

\begin{rem}
	For estimates on $\underline{N}$ associated to $\alpha$, the reader is referred to \eg \cite{Gruene2010,Worthmann2016,Gruene2017}. 
\end{rem}

Due to the monotonicity property of $V_{N}(x)$, the following is a consequence of \eqref{eq:RDP-FH-plain}:
\begin{prp}[\hspace*{-2pt}\cite{Gruene2008,Lincoln2006}]
	\label{prp:RDP}
	If \eqref{eq:RDP-FH-plain} holds for all $x \in \Omega \subset \R^n$, then
	\begin{align} \label{eq:subopt-estim}
		\alpha V_{\infty}(x) \leq \alpha J_{\infty}(x,\kappa_N(\cdot)) \leq V_{\infty}(x) 
	\end{align}
	then for all $x \in \Omega$.
\end{prp}

Inequality \eqref{eq:subopt-estim} assesses the suboptimality of the FH MPC law $\kappa_N$ \wrt the optimal infinite-horizon (IH) cost $V_{\infty}$ for any initial state $x^o \in \Omega$. 

\begin{rem}
	Variants of \eqref{eq:RDP-FH-plain} may also lead to \eqref{eq:subopt-estim}. 
	For example, \cite{Lu2020} introduce an self-triggering mechanism on the control optimization exploiting a trade-off between the necessary use of an upper bound on the optimal FH cost in the right-hand side of \eqref{eq:RDP-FH-plain} and a so-called slack variable, acting on the left-hand side of \eqref{eq:RDP-FH-plain}. 
\end{rem}

Aside from \eqref{eq:subopt-estim}, other representations of the performance loss are available in the literature (see Rem.~\ref{rem:Granzotto-estim} below).

\subsection{NMPC with terminal cost}
\label{subsec:NMPC-term-cost}

Throughout this work, consider the FH cost
\begin{align} \label{eq:FH-cost-term-cost}
	\mathcal{J}_N(x,i) = \sum_{k=0}^{N-1} l(x_u(k;x),u(k;x)) + F^i(x_u(N;x)),
\end{align}
where $\{F^i(\cdot)\}_{i \in \N_0}$, with $F^i: \R^n \ra \R_{\geq 0}$ for all $i \in \N_0$, is a sequence of \pd terminal costs. 
The optimal FH cost associated to \eqref{eq:FH-cost-term-cost} is given by 
\begin{align} \label{eq:min-MPC-cost-var}
	\mathcal{V}_N(x,i) := \inf_{u(\cdot;x) \in \R^{m \times N}} \, \mathcal{J}_N(x,i).
\end{align} 
The first element of the optimizing sequence $u^\ast(\cdot;x,i)$ defines the implicit MPC law $\kappa_N(x,i) \coloneqq u^\ast(0;x,i)$.

\begin{rem}
	In the literature, there are several available settings associated with $\mathcal{J}_N$, that incorporate mostly static terminal costs $F := F^{i} \equiv F^{i+1} $. 
	Particular examples include
	\begin{enumerate}
		\item[a)] $F$ being a local Lyapunov function, satisfying $\min_{u \in \U \subset \R^m} \{F(f(x,u)) - F(x) + l(x,u) <0 \}$ for all $x \neq 0$, \cite{Limon2006,Hu2002};
		
		\item[b)] $F= \omega \, l^\ast(x)$, for some $\omega \geq 1$, \cite[Chap. 10.2]{Gruene2017};
		
		\item[c)] $F$ as a local approximation of the optimal IH cost via a finite series, constructed by
		
		\begin{enumerate}
			\item[i)] series expansion of the dynamics, the optimal IH cost and controller $f,V_{\infty},\mu_{\infty}$ around the origin, assuming sufficient smoothness (\cite{Lucia2015a,Krener2019} and the notion of Al'brekht's method);
			
			\item[ii)] extension of $J_N$ from $x_u(N;x)$ onwards by $M$ utilities $l(x_{\kappa_{\text{loc}}}(k;x),\kappa_{\text{loc}}( x_{\kappa_{\text{loc}}}(k;x)) )$ under some known locally exponentially stabilizing control $\kappa_{\text{loc}}(\cdot)$, \cite{Koehler2021}. 
			
		\end{enumerate}
	\end{enumerate}
	Control variants of the form a) usually incorporate terminal set constraints $x_u(N;x) \in \Omega_f \subset \R^n$ while unconstrained settings amount to assuring asymptotic stability by sufficiently long prediction horizons. 
\end{rem}

A critical point in using a terminal cost function is expressed in the fact that if $V_{\infty}$ is overestimated through $\mathcal{V}_N$, $\alpha$ loses its qualitative property of describing the performance suboptimality as per \eqref{eq:subopt-estim} if no further extension is made. 
This issue may be circumvented, however, in weighting the IH optimal cost on the right-hand side of \eqref{eq:subopt-estim}, affecting the value of $\alpha$ as described in \eg \cite{Koehler2021}. 

To utilize Prop.~\ref{prp:RDP}, an adaptive dynamic programming (ADP) based method is introduced that gives rise to a sequence of terminal costs $\{F^i(\cdot)\}_{i \in \N_0}$ satisfying the aforementioned requirement. 
To be more precise, a formal iteration scheme called multi-step look-ahead value iteration (MLVI) is introduced that allows to relate the FH cost to the IH as $\mathcal{V}_N(\cdot,i) \leq V_{\infty}(\cdot)$ which in turn allows using \eqref{eq:subopt-estim} from Prop.~\ref{prp:RDP}.

\subsection{Multi-step look-ahead value iteration}

Value iteration (VI) is a particular realization of iteratively approximating the optimal IH cost. 
Therein, $V_{\infty}(\cdot)$ is approximated from below by a sequence of \pd functions $\{\hat{V}^i(\cdot)\}$, with $\hat{V}^i:\R^n \ra \R_{\geq 0}$ and $\hat{V}^i(\cdot) \leq \hat{V}^{i+1}(\cdot)$ for all $i \in \N_0$, starting at some initial guess $\hat{V}^0$ \cite{Al-Tamimi2008-VI-ADP,Heydari2014a}. 
The approximation update is based on the so-called temporal difference which is related closely to the Bellman equation and whose square may roughly be interpreted as a metric on the approximative value \wrt $V_{\infty}$. 

In recent years, so-called multi-step variants have emerged triggering an approximation update using state-input data over horizons $N>1$ \cite{Heydari2015,Luo2019}. 

As this work focuses on a predictive control framework, the following MLVI variant is introduced:

Select $\hat{V}^0(\cdot)$ \eg $\hat{V}^0(x) \equiv 0$, for iteration step $i =0$ and at each $i \in \N_0$ proceed via
\begin{enumerate}
	\item[i)] control update 
	\begin{align} \label{eq:MLVI-ctrl-update}
		&\hat{\mu}^{i}(x) = \\
		&\arginf_{u(\cdot;x)} \; \sum_{k=0}^{N-1} l(x_u(k;x),u(k;x)) + \hat{V}^i\left(x_u(N;x)\right), \nonumber
	\end{align}
	
	\item[ii)] cost evaluation 
	\begin{align} \label{eq:MLVI-cost-eval}
		\begin{split}
			&\hat{V}^{i+1}(x) = \\
			&\sum_{k=0}^{N-1} l(x_{\hat{\mu}^i}(k;x),\hat{\mu}^i(x_{\hat{\mu}^i}(k;x))) + \hat{V}^i\left(x_{\hat{\mu}^i}(N;x)\right),
		\end{split}
	\end{align} 
\end{enumerate}
and set $i = i+1$, for all $x \in \Omega$ in some set of interest $\Omega \subset \R^n$ until $i \ra \infty$, where $\hat{\mu}$ denotes the approximate IH controller. 

The following presents an intermediate result which may be consulted for terminal cost adaptation:
\begin{prp} \label{prp:MLVI}
	Let the sequence $\{\hat{V}^i(x)\}_{i \in \N_0}$ be determined via \eqref{eq:MLVI-ctrl-update}--\eqref{eq:MLVI-cost-eval}. 
	For all $N \in \N$, if for $i = 0$ and all $x \in \Omega \subset \R^n$,
	\begin{align} \label{eq:MLVI-conv-cond-init}
		\begin{split}
			\hat{V}^{0}(x)  \leq  \inf_{u(\cdot;x)} \; \sum_{k=0}^{N-1} l(x_u(k;x),u(k;x)) + \hat{V}^0\left(x_u(N;x)\right),
		\end{split}
	\end{align}
	then
	for all $i \in \N_0$ and all $x \in \Omega$,
	\begin{align} \label{eq:MLVI-iter-relat}
		\begin{split}
			\hat{V}^{i+1}(x) &= \inf_{u(\cdot;x)} \; \sum_{k=0}^{N-1} l(x_u(k;x),u(k;x)) + \hat{V}^i\left(x_u(N;x)\right) \\
			&\geq \hat{V}^{i}(x).
		\end{split}
	\end{align}
	. 
\end{prp}

\begin{IEEEproof}
	(Base case) At $i=0$, with \eqref{eq:MLVI-conv-cond-init}, we immediately obtain
	\begin{align*} 
		\hat{V}^{1}(x) &= \inf_{u(\cdot;x)} \, \sum_{k=0}^{N-1} l(x_u(k;x),u(k;x)) + \hat{V}^0\left(x_u(N;x)\right) \\
		&\geq \hat{V}^{0}(x) .
	\end{align*} 
	
	(Inductive step) Then, assuming for any $i \in \N_0$ it holds that 
	\begin{align*}
		\hat{V}^{i+1}(x) &= \inf_{u(\cdot;x)} \; \sum_{k=0}^{N-1} l(x_u(k;x),u(k;x)) + \hat{V}^{i}\left(x_u(N;x)\right) \\
		&\geq \hat{V}^{i}(x)
	\end{align*}
	it follows that
	\begin{align*}
		&\hat{V}^{i+1}(x) = \inf_{u(\cdot;x)} \, \sum_{k=0}^{N-1} l(x_u(k;x),u(k;x)) + \hat{V}^{i}(x_u(N;x)) \\
		\leq &\inf_{u(\cdot;x)} \hspace{-2pt}  \sum_{k=0}^{N-1} l(x_u(k;x),u(k;x)) + \hat{V}^{i+1}(x_u(N;x)) 
		\hspace{-1pt}= \hspace{-1pt}\hat{V}^{i+2}(x)
	\end{align*}
	as the preceding inequality holds over all $x \in \Omega$ (and thus at $x_u(N;x)$).  
	
	Hence, 
	\begin{align*}
		\hat{V}^{i+2}(x) &= \inf_{u(\cdot;x)} \; \sum_{k=0}^{N-1} l(x_u(k;x),u(k;x)) + \hat{V}^{i+1}\left(x_u(N;x)\right) \\
		&\geq \hat{V}^{i+1}(x).
	\end{align*} 
\end{IEEEproof}

\begin{rem}
	In the limit $i \ra \infty$, Proposition \ref{prp:MLVI} gives
	\begin{align*}
		\hat{V}^{\infty}(x) &= \inf_{u(\cdot;x)} \; \sum_{k=0}^{N-1} l(x_u(k;x),u(k;x)) + \hat{V}^{\infty}\left(x_u(N;x)\right) \\
		&\geq \hat{V}^{\infty}(x).
	\end{align*}
	Under some conditions (see \eg \cite[Prop.~2]{Bertsekas2017}) upper boundedness and convergence $\lim_{i \ra \infty} \hat{V}^{i}(x) = V_{\infty}(x)$, $\forall x \in \Omega$, can be concluded.
\end{rem}

\begin{rem} \label{rem:Granzotto-estim}
	Since infinite iterations are not tractable, (ML)VI is stopped at a pre-defined threshold which is commonly of the form $\hat{V}^i(x) - \hat{V}^{i-1}(x) \leq c_{\text{stop}}(\eps,x)$, $c_{\text{stop}}(\eps,x) \geq 0$. 
	For VI, \cite{Granzotto2020} provides an ``additive'' description of the performance loss in the form of $J_{\infty}(x,\hat{\mu}^{i}(\cdot)) \leq V_{\infty}(x) + \omega_{\eps}\sigma(x) $, where $\omega_{\eps} \geq 0$ and $\sigma:\R^n \ra \R_{\geq 0}$ are associated with $c_{\text{stop}}$. 
\end{rem}

Regarding stability, consider the following consequence of MLVI and sufficiently long prediction horizon:
\begin{prp} \label{prp:MLVI-LF-pair}
	Let Asm.~\ref{asm:track-utility} hold with $ N \geq \underline{N} \in \N$ for a given $\alpha \in (0,1]$ and be $F^{i}(\cdot)  = \hat{V}^{i}(\cdot)$ for any $i \in \N_0$.  
	Then, for any $i \in \N_0$, $\mathcal{V}_N(\cdot,i)$ decays with at least $\alpha \cdot \alpha_l(|\cdot|)$.
\end{prp}
\begin{IEEEproof}
	Note that under MLVI $\mathcal{V}_N(x,i) \equiv V_{i \cdot N}(x) \geq V_{N}(x)$. 
	By assumption, \eqref{eq:RDP-FH-plain} holds for $\alpha$ with horizon $M = i \cdot N \geq \underline{N}$. 
\end{IEEEproof}

Compared to the time-invariant setting, a suboptimality analysis of time-variant schemes appears to be more complex. 
The main reason is that offline computed bounds, such as \eg a $\mathcal{KL}$\ \footnote{A function $\beta:\R_{\geq 0} \times \R_{\geq 0} \ra \R_{\geq 0}$ belongs to class $\mathcal{KL}$ if for fixed $s$, $\beta(r,s)$ is of class $\mathcal{K}$ and for fixed $r$, $\beta(r,s)$ is decreasing \wrt $s$ and $\lim_{s \ra \infty} = 0$.} function associated with an asymptotic (or exponential) controllability assumption \cite{Gruene2009}, may, in general, not hold during the transient behaviour. 
Even though a related analysis of time-varying MPC (refer to \eg \cite[Asm.~6.30 and Chap.~10.2]{Gruene2017}) is possible \emph{if} the time evolution is known, the available approaches utilize a uniform bound (over time) and thus the resulting performance estimate may be conservative. 
Another issue arises due to the following: MLVI iterates the approximation $\hat{V}$ over the entire space $\Omega$ with exact update \ie equations \eqref{eq:MLVI-ctrl-update}--\eqref{eq:MLVI-cost-eval} can be solved exactly. 
Parametric approximation structures are therefore introduced to cope with the computational complexity and (local) tractability of the iteration scheme (see Section \ref{subsec:approx-alg}).

\section{Main Results}

In the following, it is assumed that $F^0(x) \equiv 0$ \ie the controller at initial time $t=0$, $\kappa_N(x,0)$, is obtained via FH optimization without terminal cost. 
Therefore, if the terminal cost adaptation follows the MLVI procedure, the closed-loop IH cost suboptimality is at least $\alpha$ associated to $\underline{N}$ according to Prop.~\ref{prp:MLVI-LF-pair}. 
Yet, as the terminal cost approaches the optimal IH cost, controller performance improvement can be expected. 
As briefly mentioned in Subsec. \ref{subsec:NMPC-term-cost}, utilizing offline computed local approximations $\hat{V}^i$ to $V_{\infty}$ in $\eqref{eq:FH-cost-term-cost}$ requires ingredients such as \eg a local controller and its region of attraction or $r-$differentiability of system, controller and $V_{\infty}$. 
The former may be difficult in case the linearized dynamics around the desired set point $x= 0$ are not stabilizable or the input constraint set is $\U \neq \R^m$. 

Online cost adaptation, through which time-variance is induced, offers an additional degree of freedom that may counteract some drawbacks of offline designs at the (unavoidable) expense of shifting computational load to the online procedure. 
A particular feature and focus of this work lies in the increase and (online) estimation of the performance bound in Prop.~\ref{prp:RDP}. 
\wrt to an initial value. 
However, in case the terminal cost is adapted online using only local state-input information, \eqref{eq:subopt-estim} in Prop.~\ref{prp:RDP} is hard to establish beforehand as it depends on the trajectory chosen by the controller.  
In this section, two variants of a performance estimation as per \eqref{eq:subopt-estim} are introduced which relate the point-wise relaxed decay to a global, IH statement. 
Then, to provide an algorithm, a parametrized MLVI is chosen as a strategy for terminal cost updating. 

\subsection{Offline a priori performance bound}

Let $x(t)$ be the closed-loop trajectory associated to the feedback $u(t) \coloneqq \kappa_N(x(t),t)$, which generates the utility $l(t) \coloneqq l(x(t),u(t))$. 

The following assigns an overall performance to a time-variant implicit control law $\kappa_N$ if decay of specific form is given: 

\begin{thm} \label{prp:subopt-estim-geometric}
	Let Asm.~\ref{asm:track-utility} hold and be $\underline{N} \in \N$ for a given $\alpha \in (0,1]$ such that \eqref{eq:RDP-FH-plain} for $N \geq \underline{N}$. 
	Let $N \geq \underline{N}$ as well as $\mathcal{J}_N(x,0) \equiv J_N(x)$. 
	If there exists a \pd function $\bar{V}_N: \R^n \times \N_0 \ra \R_{\geq 0}$ satisfying $\bar{V}_N(x,t) \geq \mathcal{V}_N(x,t)$, where $\mathcal{V}_N$ is from \eqref{eq:min-MPC-cost-var}, and for all $(x(t),t) \in \R^n \times \N_0$, 
	\begin{align} \label{eq:prp-rel-decay-geom}
		\mathcal{V}_N(x(t),t) \geq \bar{V}(x(t+1),t+1) + \alpha c_t + e(t), 
	\end{align}
	where 
	\begin{align}
		c_t = \sum_{k=0}^{t} \gamma^k l(t-k), \quad \gamma \in (0,1-\alpha)
	\end{align}
	and
	\begin{align}
		e(t) = \begin{cases}
			&\hspace*{-1em} e(t-1) + \bar{V}_N(x(t),t) - \bar{V}_N(x(t-1),t-1) \\
			&+ \alpha \sum_{p=0}^{t-1} \gamma^p l(t-p), \; t \geq 2, \\[0.5em]
			& \hspace*{-1em} \bar{V}_N(x(1),1) - \mathcal{V}_N(x(0),0) + \alpha \gamma l(0), \; t=1 \\[0.5em]
			&\hspace*{-1em} 0, \; t = 0,
		\end{cases}
	\end{align}
	then the origin is asymptotically stabilized for all $x^o \in \R^n$ and
	\begin{align}
		\frac{\alpha}{1-\gamma} J_{\infty}(x^o,u(\cdot)) \leq V_{\infty}(x^o).
	\end{align}
\end{thm}

\begin{IEEEproof}
	Subsequently, a distinct performance estimate related to the initial suboptimal decay is presented establishing $\mathcal{V}$ as a Lyapunov function.
	
	Consider $e$ at time $n \in \N$, $n \geq 2$, which gives
	\begin{align*}
		e(t) &= e(t-1) + \bar{V}_N(x(t),t) \\ 
		&\hspace*{-3pt}-\bar{V}_N(x(t-1),t-1) + \alpha \sum_{p=0}^{t-1} \gamma^p l(t-1-p) \\
		&\hspace*{-12pt}= e(t-2) + \bar{V}_N(x(t),t) - \bar{V}_N(x(t-2),t-2)  \\
		&\hspace*{-3pt}+\sum_{k=t-2}^{t-1} \alpha \sum_{p=0}^{k} \gamma^p l(k-p) \\
		&\hspace*{-12pt}= \bar{V}_N(x(t),t) - \mathcal{V}_N(x^o,0) + \alpha \sum_{k=0}^{t-1} \sum_{p=0}^{k} \gamma^p l(k-p),
	\end{align*}
	where the second equality is obtained by dissolving $e(t-1)$. 
	
	It follows that
	\begin{align*}
		0 \leq  \,\alpha \sum_{k=0}^{t-1} c_k 
		=  \,e(t) - \bar{V}_N(x(t),t) + \mathcal{V}_N(x^o,0),
	\end{align*}
	which also holds for $t=1$.

	Furthermore, from \eqref{eq:prp-rel-decay-geom}, $e(t) \leq \mathcal{V}_N(x(t),t)-\bar{V}_N(x(t+1),t+1)$. 
	Together with the above equation this yields
	\begin{align*}
		\alpha \sum_{k=0}^{t-1} c_k 
		\leq \; &\mathcal{V}_N(x^o,0) - \bar{V}_N(x(t+1),t+1) \\ 
		&- \bar{V}_N(x(t),t) + \mathcal{V}_N(x(t),t) \\
		\leq \;&\mathcal{V}_N(x^o,0) - \bar{V}_N(x(t+1),t+1).
	\end{align*}
	
	The second part of the proof considers the limiting case and the resulting suboptimality.  
	
	By extension of the previous inequality one obtains for $t \ra \infty$,
	\begin{align} \label{eq:decay-limit-UB-IHcost}
		\begin{split}
			\alpha \sum_{t=0}^{\infty} c_t& \leq \mathcal{V}_N(x^o,0) - \mathcal{V}_N(x(t),t)\\
			&\leq \mathcal{V}_N(x^o,0) = V_N(x^o) \leq V_{\infty}(x^o). 
		\end{split}
	\end{align}
	Given the decay rate of \eqref{eq:prp-rel-decay-geom}, $\mathcal{V}_N$ is a valid Lyapunov function if there exists uniform bounds. 
	By assumption, using the above inequality, $\mathcal{V}_N(x, \cdot)$ is lower and upper bounded by $\alpha_l$ and $\alpha_V$, respectively. 
	Therefore, $\lim_{t \ra \infty} x(t) =0$ follows. 
	
	Then, by the Cauchy product for convergent series, 
	\begin{align*}
		\alpha \sum_{t=0}^{\infty} c_t = \alpha \sum_{i=0}^{\infty} \gamma^i \sum_{j=0}^{\infty} l(j) = \frac{\alpha}{1-\gamma} J_{\infty}(x^o,u(\cdot)), 
	\end{align*}
	which completes the proof.
\end{IEEEproof}

It should be noted that computing $\underline{N}$ for $\Omega = R^n$ can be intractable and subsets of the state space should be investigated \ie those on which \eqref{eq:prp-rel-decay-geom} holds and $\Omega$ is positive invariant. 
This observation also affects the inclusion of state constraints which requires further research. 

\begin{rem}
	The above result in Thm.~\ref{prp:subopt-estim-geometric} is a modified and repurposed variant of the self-triggered MPC approach \cite{Lu2020}, allowing for slack in the required decay and thereby reducing the effect of using $\bar{V}_N$ in the decay estimation. 
\end{rem}
Somewhat similar to \cite[Rem.~7]{Lu2020}, the true decay may deviate from the desirable one using $\gamma$. 
To cope with that fact, the slack variable is adjusted in the following.

\subsection{Online performance bound estimation}

The aforementioned analysis provides a performance bound based the decay of $\mathcal{V}_N$ related to $\gamma$.
Note that the true decay can not be adjusted per se but is a result of terminal cost modification. 
Also note that the slack variable is a fixed construct. 
Therefore, in the following a lower bound estimation on $\alpha$ of \eqref{eq:subopt-estim} is proposed that aims to adjust the suboptimality estimate based on current information:

\begin{thm} \label{prp:subopt-estim-telescop}
	Let Asm.~\ref{asm:track-utility} hold and be $\underline{N} \in \N$ for a given $\alpha \in (0,1]$ such that \eqref{eq:RDP-FH-plain} for $N \geq \underline{N}$. 
	Let $N \geq \underline{N}$ as well as $\mathcal{J}_N(x,0) \equiv J_N(x)$. 
	If there exists a \pd function $\bar{V}_N: \R^n \times \N_0 \ra \R_{\geq 0}$ and a convergent sequence $\{s_k\}_{k \in \N_0}$ satisfying $\bar{V}_N(x,t) \geq \mathcal{V}_N(x,t)$, $s_k >0$ for all $k \in \N_0$ and 
	\begin{align} \label{eq:prp-rel-decay-telecop}
		\mathcal{V}_N(x(t),t) \geq \bar{V}(x(t+1),t+1) +  \sum_{k=0}^{t} s_k l(t-k) 
	\end{align}
	for all $(x(t),t) \in \R^n \times \N_0$, where $s_k = a_k - a_{k+1} + b_k$, $\{a_k\}_{k \in \N_0},\{b_k\}_{k \in \N_0}$ convergent with $\lim\limits_{k \ra \infty} a_k = a_{\infty}$ and $\mathcal{V}_N$ is from \eqref{eq:min-MPC-cost-var}, then
	\begin{align} \label{eq:IH-cost-telescopic}
		\left( a_0 - a_{\infty} + \sum_{t=0}^{\infty} b_t \right) J_{\infty}(x^o) \leq V_{\infty}(x^o)
	\end{align}
	and $x=0$ is asymptotically stabilized for all $x^o \in \R^n$. 
\end{thm}

\begin{IEEEproof}
	The proof differs to that of Thm. \ref{prp:subopt-estim-geometric} marginally. 
	In the limit, it is used that
	\begin{align*}
		&\sum_{t=0}^{\infty} \sum_{k=0}^{t} s_k l(t-k) = \sum_{i=0}^{\infty} s_i \sum_{j=0}^{\infty} l(j) \\
		= &  \left( \sum_{t=0}^{\infty} (a_t - a_{t+1}) + \sum_{t=0}^{\infty} b_t  \right) J_{\infty}(x^o,u(\cdot)),
	\end{align*}
	which, after employing the telescopic sum property, gives the result.  
\end{IEEEproof}

The above estimate extends the previous established a~priori computation on the following point: 
Given a sequence $\{a_k\}_{k \in \N_0}$, which may take on the form $a_k - a_{k+1} = \gamma^k$, $b_k$ takes the role of the slack variable, analogous to $e$. 
Yet, while $e$ is fixed, $b_k$ is not per se allocated but computed through decay experience a posteriori to the terminal cost update (see Alg.~\ref{alg:mlvi-mpc} below). 
That is, knowing $a_0 - a_{\infty}$ (by user design of specific $\{a_k\}$), assigning and accumulating $b_k$ online provides an estimate on the expected overall performance. 

\begin{rem} \label{rem:seq-a}
	The sequence $\{a_k\}_{k \in \N_0}$ can be chosen such that $a_0 - a_{\infty} \in (0,1)$ while $b_k$ accumulates the error and overshoot provided by $\mathcal{V}_N$ relative to the desired decay. 
	For example, take $a_{k+1} = a_k -\frac{\bar{a}}{k(k+1)}$ for some $\bar{a} \in (0,1)$ which gives $\sum_{k=1}^{\infty} (a_k- a_{k+1}) = \bar{a}$. 
	For $t=0$, $s_0 = \bar{\alpha}(0) + \beta_0$ is assigned, where $\bar{\alpha}(t)$ is an approximation on the relaxed decay rate that can be extracted at time step $t$ via $\bar{\alpha}(t) = \mathcal{V}_N(x(t),t) - \bar{V}_N(x(t+1),t)/l(t)$ (for nonzero $l(t)$). 
	Then the suboptimality bound $\alpha$ per Prop.~\ref{prp:RDP} is
	\begin{align} \label{eq:estimate-a-seq}
		\alpha = \bar{\alpha}(0) + \bar{a} + \sum_{t=0}^{\infty} \beta_t.
	\end{align}
	As a special case of the above take $\bar{a}=0$ (or analogously $a_0 = \bar{\alpha}(0)$ and $a_k =0$ for all $k \geq 1$ in \eqref{eq:IH-cost-telescopic}) so that the accumulated $\beta_t$ does evaluate the improvement \wrt the initial performance. 
\end{rem}

\begin{rem}
	Note that for stability, based on the estimate $\bar{\alpha}(t)$, the terminal cost increment is upper bounded such that decay of the optimal FH cost is preserved. 
	Specifically, using \eg $\bar{V}_N(x^+,t) := \mathcal{J}_N(x^+,\tilde{u}(\cdot),t)$ under an extended control sequence of the form $\tilde{u}(\cdot) := \{ u^\ast(1;x), \dotsc, u^\ast(N-1;x), \bar{u} \}$ where for example $\bar{u} =0$, the terminal cost adjustment must satisfy
	\begin{align}\label{eq:term-cost-adapt-constr}
		F^{t+1}\left(x_{N+1} \right) \leq F^{t}\left(x_{N+1}\right) + \bar{\alpha}(t) l(t) ,
	\end{align}
	utilizing $x_{N+1} = f(x_{u^\ast}(N;x(t)),\bar{u})$.
\end{rem}

Hence, it can be observed that an ultimate restriction of performance improvement with the given setting comes with the relaxed decay of the current time step $t$ itself. 
However, the presented conditions exploits plain monotonicity of $V_N$ whereas modifications of \eg \eqref{eq:decay-limit-UB-IHcost} can be introduced using comparative results such as $V_N(x) \leq (1+ \eps(x) )V_{\infty}(x)$ (refer to \cite{Koehler2021}) or as in \cite{Granzotto2020} (see Rem. \ref{rem:Granzotto-estim}).
Further alternatives may include utilizing the concept of $M-$step stability \cite[Def. 6]{Noroozi2020} to soften the decay requirement. 
Specifically, the decay is guaranteed only after $M$ time step between which the adaptation mechanism collects data and adjusts the terminal cost rather freely.

\subsection{Online adaptation of terminal cost and bound estimation}
\label{subsec:approx-alg}

For terminal cost adaptation, a parametric form of MLVI is applied utilizing linear combinations of basis function (see \eg \cite{Heydari2014a}).
Commonly, the adaptive approximation function $\hat{V}$ is called \textit{critic} in the ADP literature \cite{Liu2000}. 

Let $\phi_c: \R^n \ra \R^l$, $l \in \N$, be the basis functions of the critic. 
The critic approximates the IH optimal cost as 
\begin{align} \label{eq:critic-param}
	\hat{V}^{i}(x) = \left\langle w_c^{i}, \phi_c(x)  \right\rangle  = \sum_{j=1}^{l} w_{c,j}^{i} \phi_{c,j}(x) ,
\end{align}
where $\{w_c^i\}_{i \in \N_0}$ is a sequence of critic weights to be found. 
Given \eqref{eq:critic-param}, an online MLVI based update $w_c^t \mapsto w_c^{t+1}$ is proposed as follows: consider \eqref{eq:MLVI-cost-eval} which in parametrized form locally at time step $t$ reads as  
\begin{align}
	&\left\langle w_c^{t+1}, \phi_c(x(t)) \right\rangle =  \\
	&\underbrace{\sum_{k=0}^{N-1} l (x_{u^\ast}(k;x(t)),u^\ast(k;x(t))) + \left\langle w_c^{t}, \phi_c( x_{u^\ast}(N;x(t)) ) \right\rangle}_{=: \beta(x(t),w_c^t)}. \nonumber
\end{align}  
The above represents an underdetermined system on $l$ weights to which a local solution $\hat{w}_c^{t+1}$ at $x$ can be found via \eg recursive least squares or gradient descent. 
For example, iterate through
\begin{align} \label{eq:weight-GD}
	\begin{split}
		&w_c^{t,i+1} = \\
		&w_c^{t,i} - \eta \left(  \left\langle w_c^{t,i}, \phi_c(x(t)) \right\rangle - \beta(x(t),w_c^t) \right) \phi_c(x(t))
	\end{split}
\end{align}
until $|w_c^{t,i} - w_c^{t,i-1}  |\leq \eps_w$, $\eps_w>0$, then setting $w_c^{t,i} = \hat{w}_c^{t+1}$.
Since the terminal cost adaptation is subject to constraints \eqref{eq:term-cost-adapt-constr}, the solution may be projected onto the set via
\begin{subequations} \label{eq:critic}
	\begin{align}
		\min_{w} \quad &\left(   w -  \hat{w}_c^{t+1}  \right)^2 \\
		\begin{split}
			\text{s.t.} \quad & \langle w - w_c^t,\phi_c(x_{N+1}) \rangle \leq \bar{\alpha}(t)l(t) 
		\end{split}\\
		&0 < \langle w, \phi_c(x) \rangle, \; \forall x \neq 0. \label{eq:w-opt}
	\end{align}
\end{subequations}

A feasible solution to \eqref{eq:critic} is given by $w = w_c^t$ \ie leaving the terminal cost unchanged. 
The last inequality \eqref{eq:w-opt} assigns the new weight such that the terminal cost is \pd for all possible states (cf.~\cite{Beckenbach2020a}). 
This requirement is easily satisfied if \eg $\phi_c$ is \pd, rendering the above a convex quadratic program. 
Other approaches \eg purely gradient based, do not pose this requirement explicitly but consider suitable configuration of the update law in interaction with feedback parametrization to satisfy \eqref{eq:w-opt} implicitly (refer to \eg \cite{Vamvoudakis2010}). 

After the terminal cost update is done, the performance estimate correction $b_t $ can be computed based on the total decay of the FH cost until time $t$ related to the convoluted decay as
\begin{align} \label{eq:b_n-comp}
	\begin{split}
		&\sum_{k=0}^{t-1} \sum_{p=0}^{k}s_p l(k-p) + \sum_{p=0}^{t-1} s_p l(n-p) + (a_t - a_{t+1} + b_t)l(0)\\
		&\leq \mathcal{V}_N(x(0),0) - \bar{V}_N(x(t+1),t+1)).
	\end{split}
\end{align}
Going along \eqref{eq:prp-rel-decay-telecop}, $b_n $ is chosen as large as possible \ie equating the above -- furthermore $b_t$ can be computed at $t+1$ using $\mathcal{V}_N(x(t+1),t+1)$ to obtain a more accurate estimate. 

\begin{rem}
	Note that $\langle  w_c^{i},\phi_c(x) \rangle$ can furthermore be taken as approximant of the optimal FH cost, thus allowing to state the implicit control law $\kappa_N$ in closed-form for special cases \eg affine system and quadratic utility.
\end{rem}


Alg.~\ref{alg:mlvi-mpc} summarizes the control and adaptation cycle, henceforth denoted as MLVI-MPC, including the performance bound estimate. 
The terminal cost represents the critic, approximating the IH tail via the parametric form \eqref{eq:critic-param}. 
After the FH problem is solved at time $t$, a lower bound on the decay rate is estimated using $ \bar{V}_N$ such that the terminal cost adaptation does not harm FH cost decay. 
Then, given the updated terminal cost using weights from \eqref{eq:critic}, the correction term $b_t$ can be computed and added to previous values. 
\begin{algorithm}[h]
	\SetAlgoLined
	\textbf{Init:} horizon $N \in \N$ s.t. $\alpha >0$, basis $\phi_c:\R^n \ra \R^l$, terminal cost $F^i \equiv \hat{V}^i$, upper bound $\bar{V}_N$ \\
	\textbf{Input:} $w_c^0 = 0$, sequence $\{a_k\}$, step size $\eta>0$, stop criterion $\eps_w >0$\\
	\textbf{For each $t \in \N_0$ do:} \\
	\hspace*{4pt}1: Minimize \eqref{eq:FH-cost-term-cost} to obtain $\mathcal{V}_N(x(t),t),u^\ast(\cdot,x(t))$ \\
	\hspace*{4pt}2: Compute approximate index for $x \neq 0$ by \\ \hspace*{1.3em}$\bar{\alpha}(t) = (\mathcal{V}_N(x(t),t) - \bar{V}_N(x(t+1),t))/l(t)$ \\
	\hspace*{4pt}3: Calculate \eqref{eq:weight-GD} until $|w_c^{t,i} - w_c^{t,i-1}  |\leq \eps_w$\\
	\hspace*{4pt}4: Get $w_c^{t+1}$ from \eqref{eq:critic} and update $F^t \ra F^{t+1}$\\
	\hspace*{4pt}5: Calculate $b_t$ satisfying \eqref{eq:b_n-comp} \\
	\textbf{Output:} $\kappa_N(x(t),t)$, $w_c^{t+1}$, $\sum_{k=0}^{t} \beta_k$
	\caption{MLVI-MPC}
	\label{alg:mlvi-mpc}
\end{algorithm}


\section{Case Study}

To demonstrate the proposed algorithm and investigate its performance, the system considered in this case study is designed using converse optimality results for discrete time systems \cite{Goehrt2020} in which the optimal IH cost is known. 
Suppose
\begin{align*}
	V_{\infty}(x) = x^\top P x, \quad P = \begin{bmatrix}
		5 & 1 \\ 1 & 3
	\end{bmatrix}.
\end{align*}
Consider further dynamics of the form
\begin{align*}
	x^+ = f(x) + Bu, \quad B = \begin{bmatrix}
		0 & 1
	\end{bmatrix}^\top
\end{align*}
as well as the utility
\begin{align*}
	l(x,u) = x^\top Q x + u ^\top R u, \; Q = \begin{bmatrix}
		2 & 0 \\ 0 & 1
	\end{bmatrix}, \; R = 1.
\end{align*}
By converse optimality, the nonlinear drift is 
{\footnotesize
	\begin{align*}
		f(x) = \begin{bmatrix}
			0.4608 & -0.044 \\ -0.044 & 1.1641
		\end{bmatrix} \begin{bmatrix}
			\frac{a_1(x)}{b(x)} & -\frac{a_2(x)}{b(x)} \\ \frac{a_2(x)}{b(x)} & \frac{a_1(x)}{b(x)}
		\end{bmatrix} \begin{bmatrix}
			1.7013 & 0.3249 \\ 0.3249 & 1.3764
		\end{bmatrix} x,  \;
\end{align*} }%
utilizing $a_1(x) = 1 + x_1^2$, $a_2(x)= x_1x_2$ and $b(x) = \sqrt{a_1^2(x) + a_2^2(x)}$. 
A stabilizing prediction horizon of $N=4$ for $ x^o \in \Omega = [-4,4]^2$ has been determined heuristically for which the (vicinity of the) origin is reached after at most $N_{\text{sim}} = 14$ time steps. 
The sequence $\{a_k\}$ is chosen as per Rem. \ref{rem:seq-a} with $\bar{a}=0.3$.
Subsequently, if $\bar{\alpha}(0) + \bar{a} >1$ in \eqref{eq:estimate-a-seq}, then the accumulated $\beta_n$ must be negative. 
The terminal cost basis is chosen to be $\phi_c(x) = [x_1^2, \, x_1 x_2, \, x_2^2]^\top$. 

\begin{figure}[H]
	\centering 
	\includegraphics[width = 0.85\columnwidth]{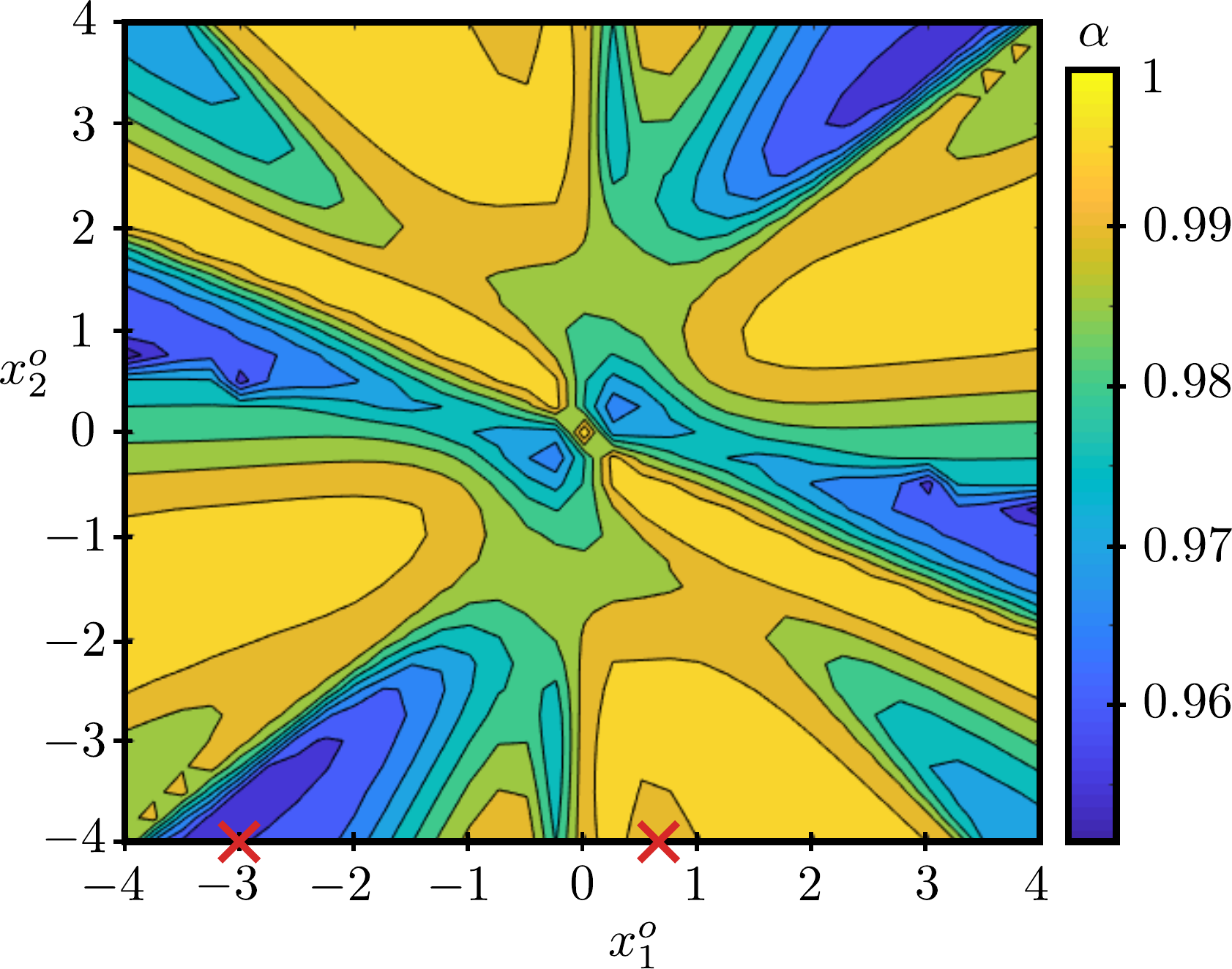}
	\caption{Cost comparison IH optimal cost vs. MLVI-MPC. It can be seen that the IH cost is close to the $V_{\infty}$ according to the suboptimality index. Regarding \eqref{eq:subopt-estim}, $\alpha = 0.95$ holds for $x^o \in \Omega$.}
	\label{fig:cost_comp_alpha}
\end{figure}

Fig.~\ref{fig:cost_comp_alpha} depicts the IH cost suboptimality $V_{\infty}(x^o) / J_{\infty}(x^o,\kappa_N(\cdot,\cdot))$ under the controller $\kappa_N$ generated through Alg.~\ref{alg:mlvi-mpc} over a state space grid $x^o \in [-4,4]^2$. 

\begin{figure*}[t]
	\centering
	\begin{subfigure}[t]{0.45\columnwidth}
		\centering
		\includegraphics[width = \columnwidth]{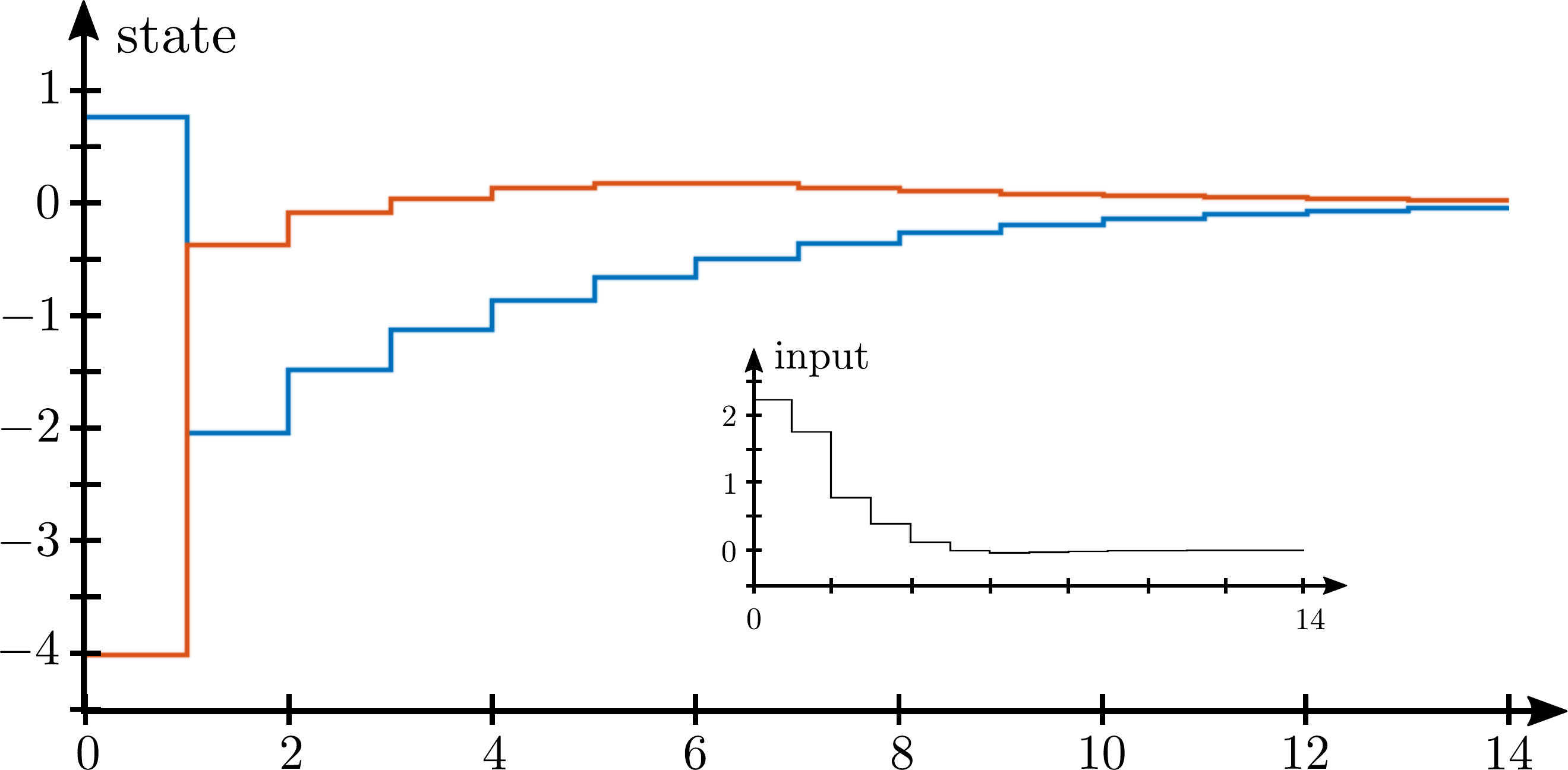}
		\caption{State and input trajectory.}
	\end{subfigure}
	\hspace{1em}
	\begin{subfigure}[t]{0.45\columnwidth}
		\centering
		\includegraphics[width = \columnwidth]{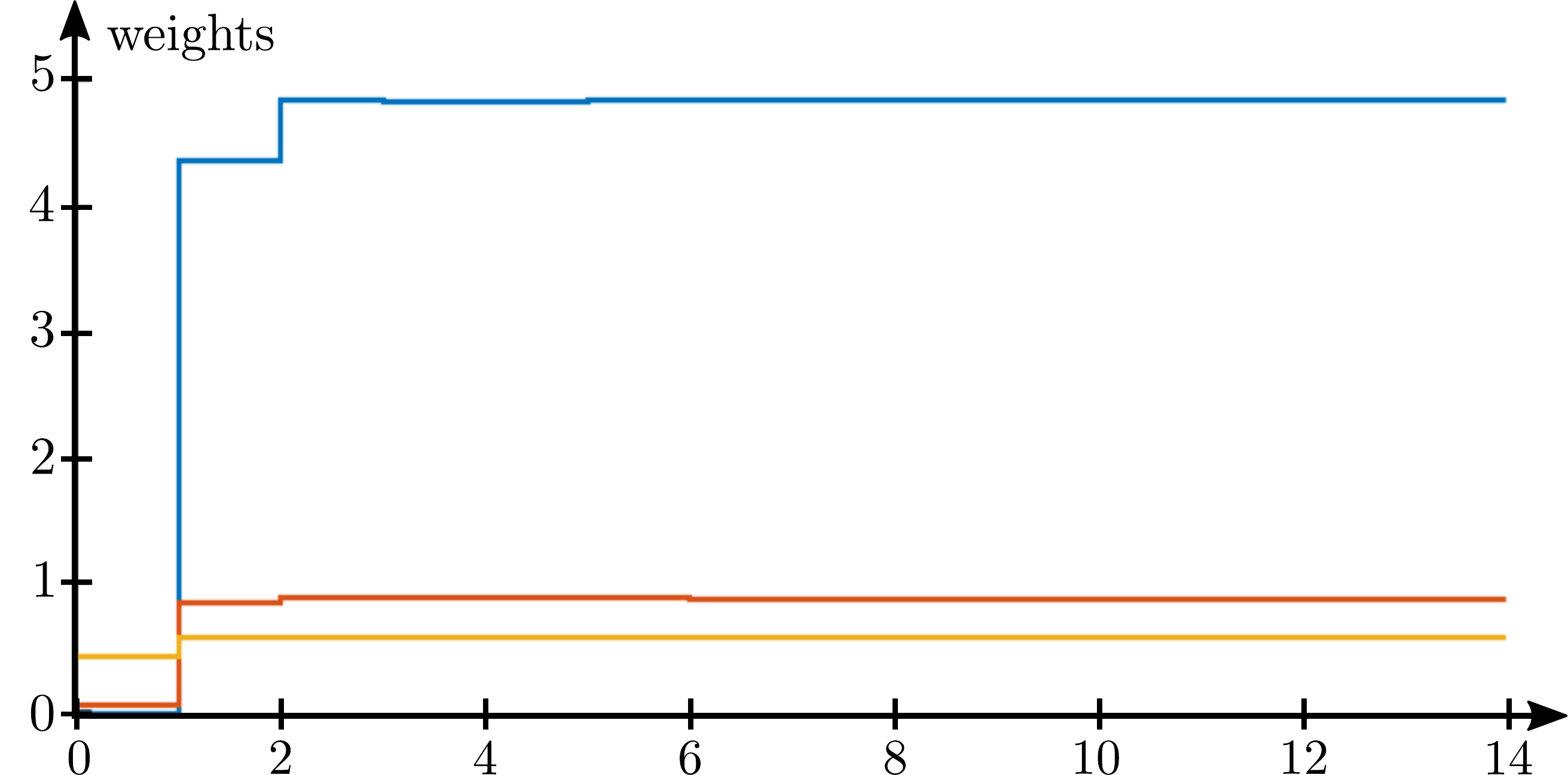}
		\caption{Weights of terminal cost.}
	\end{subfigure}
	\hspace{1em}
	\begin{subfigure}[t]{0.45\columnwidth}
		\centering
		\includegraphics[width = \columnwidth]{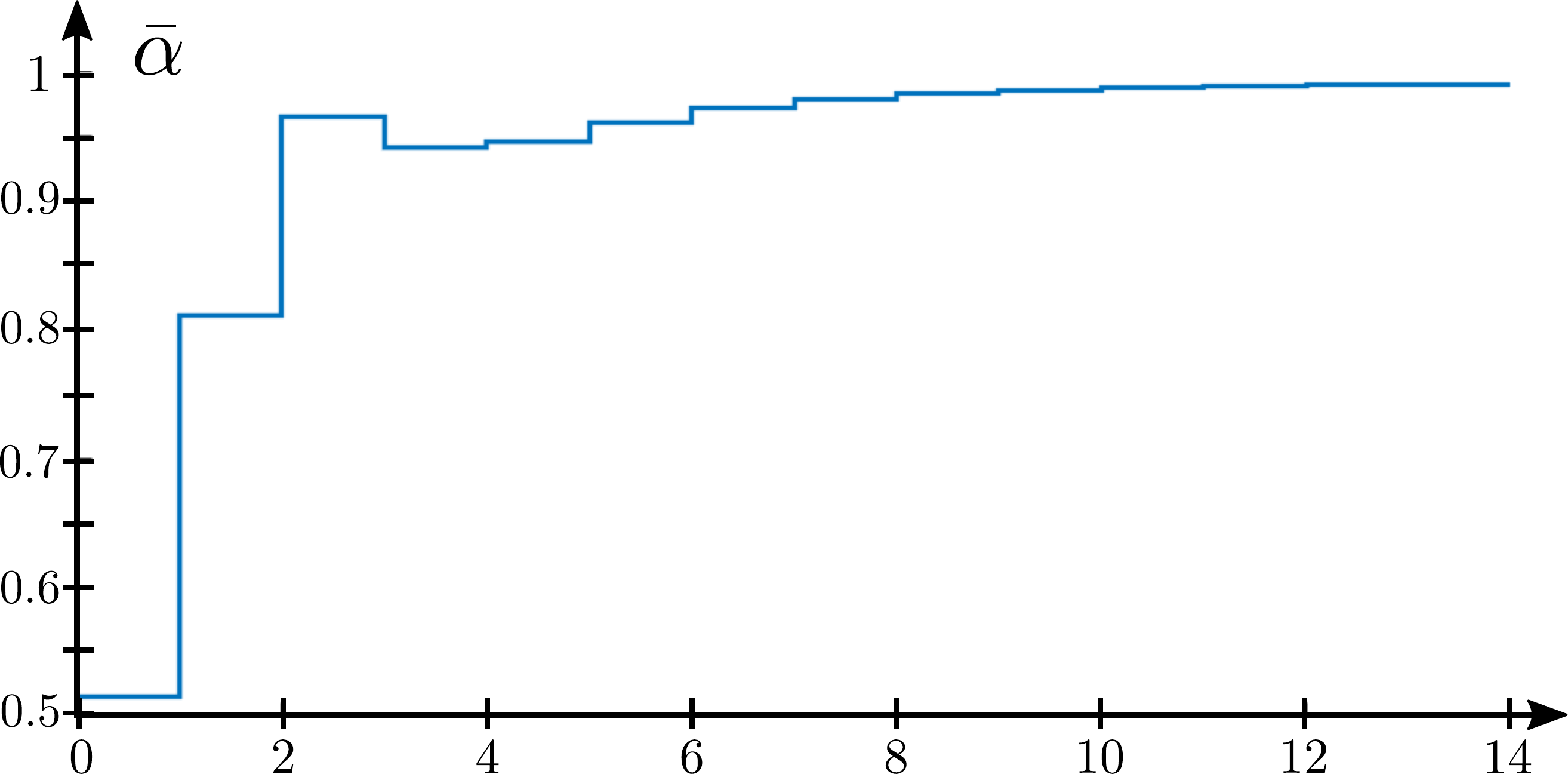}
		\caption{Online estimate on suboptimality index.}
		\label{fig:alpha-1}
	\end{subfigure}
	\hspace{1em}
	\begin{subfigure}[t]{0.45\columnwidth}
		\centering
		\includegraphics[width = \columnwidth]{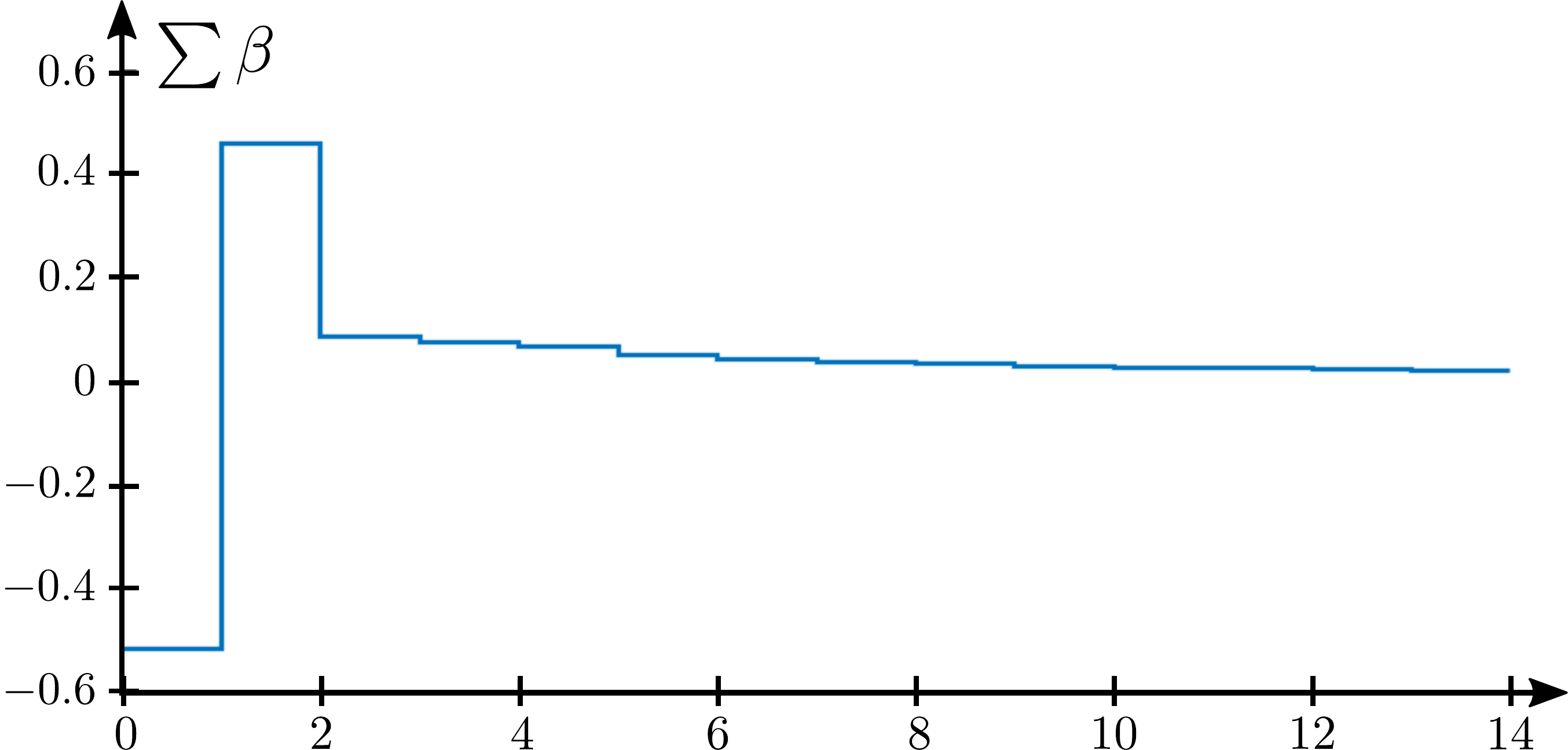}
		\caption{Summed performance correction term.}
		\label{fig:beta-summed-1}
	\end{subfigure}
	\centering
	\begin{subfigure}[t]{0.45\columnwidth}
		\centering
		\includegraphics[width = \columnwidth]{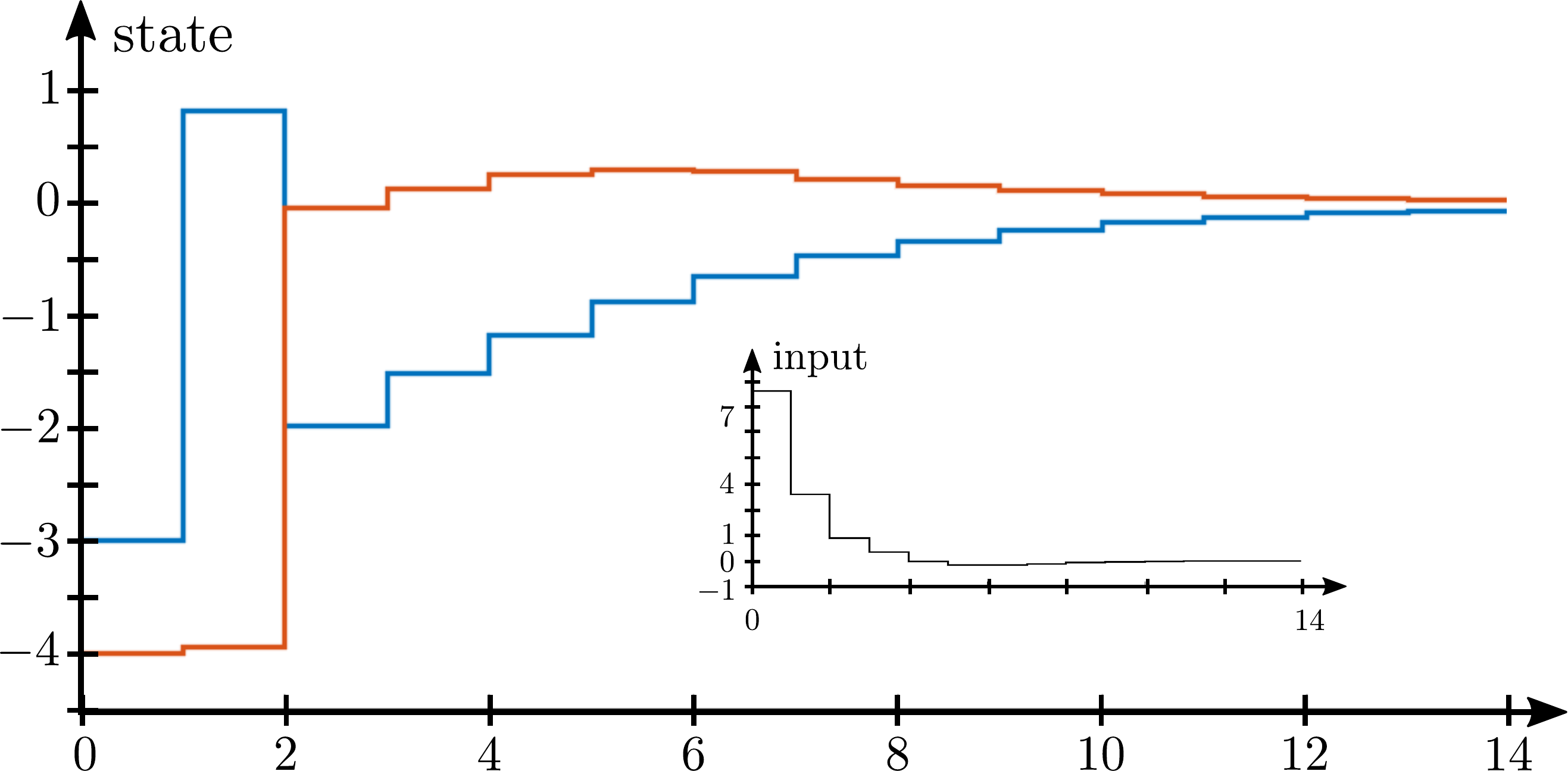}
		\caption{State and input trajectory.}
	\end{subfigure}
	\hspace{1em}
	\begin{subfigure}[t]{0.45\columnwidth}
		\centering
		\includegraphics[width = \columnwidth]{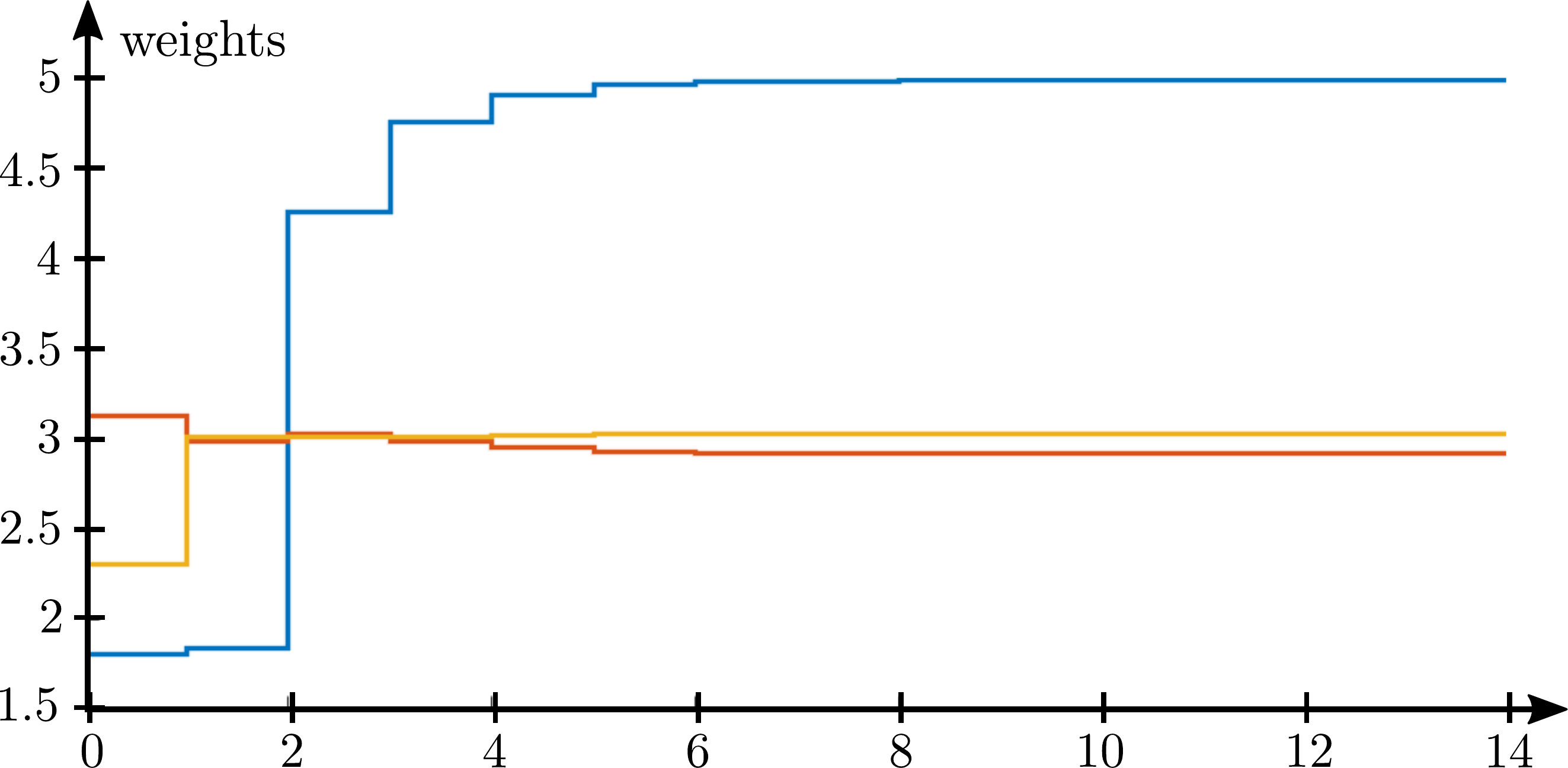}
		\caption{Weights of terminal cost.}
	\end{subfigure}
	\hspace{1em}
	\begin{subfigure}[t]{0.45\columnwidth}
		\centering
		\includegraphics[width = \columnwidth]{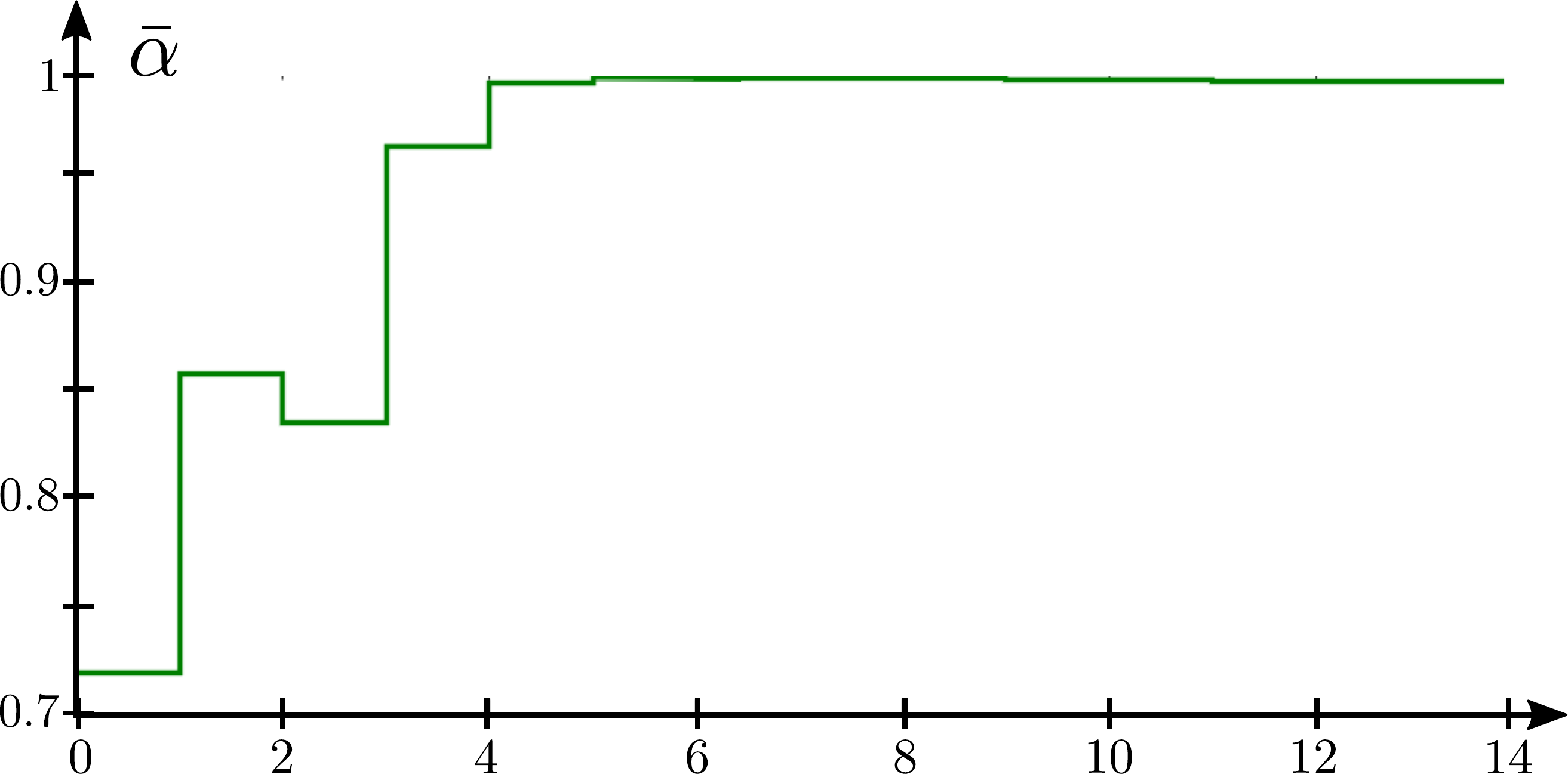}
		\caption{Online estimate on suboptimality index.}
		\label{fig:alpha-3}
	\end{subfigure}
	\hspace{1em}
	\begin{subfigure}[t]{0.45\columnwidth}
		\centering
		\includegraphics[width = \columnwidth]{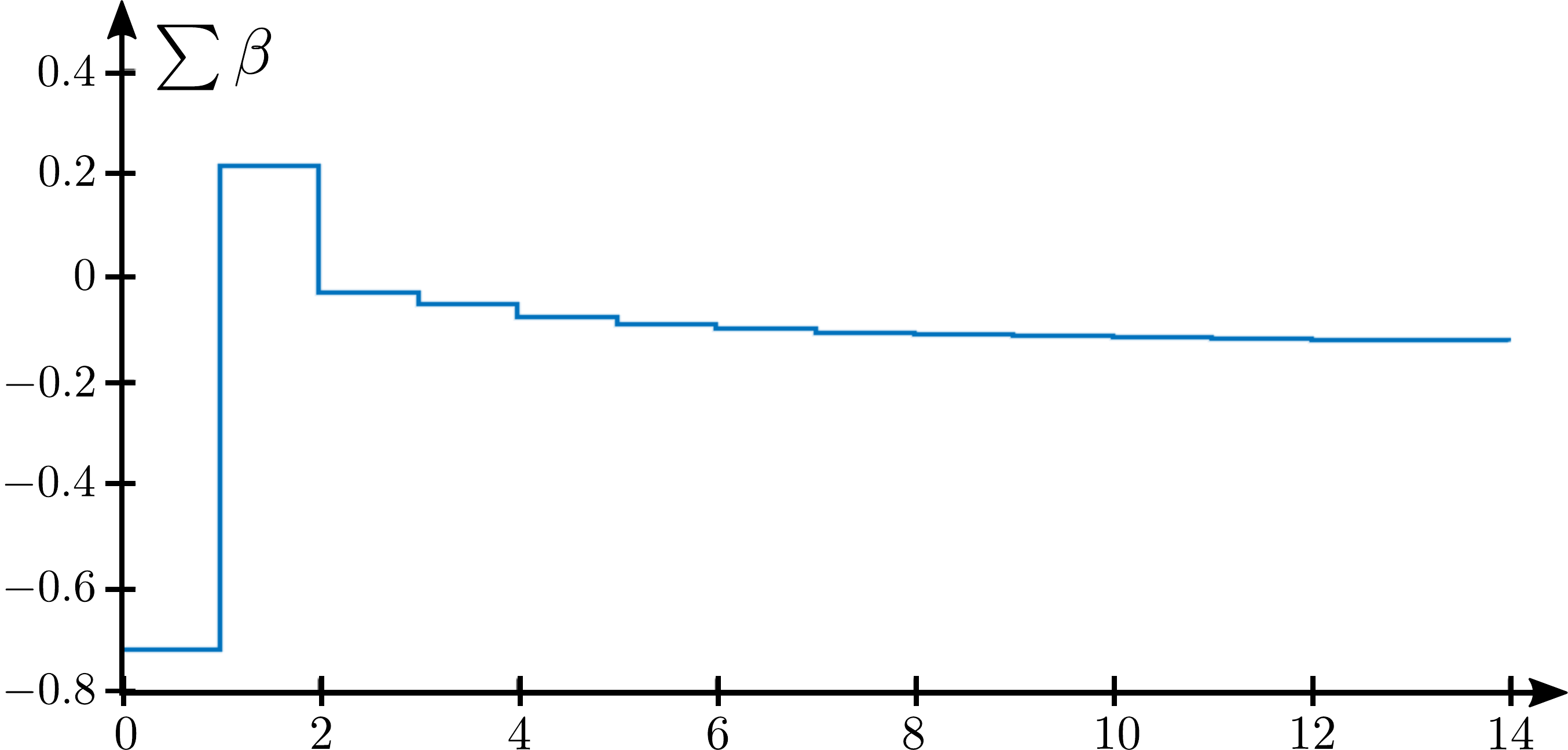}
		\caption{Summed performance correction term.}
		\label{fig:beta-summed-3}
	\end{subfigure}
	\caption{Trajectories of state (blue $x_1(t)$, red $x_2(t)$), terminal cost weights (blue $w_1(t)$, red $w_2(t)$, yellow $w_3(t)$), online suboptimality estimate $\bar{\alpha}(t) $ and the accumulation of $b_t$ for $x^o = [0.75,-4]^\top$ and $x^o = [-3,-4]^\top$.}
	\label{fig:data-samples}
\end{figure*}

Fig.~\ref{fig:data-samples} presents results for two initial conditions yielding distinct performances. 


It can be seen that while the origin is stabilized the weights of the terminal cost are adapted and the suboptimality reduces. 
Particularly concerning Fig.~\ref{fig:alpha-1}--\ref{fig:beta-summed-1}, with the initial estimate $\bar{\alpha}(0) = 0.51$, the estimated total performance needs not to be corrected downwards as can be seen in the terminal accumulated correction $\sum_{t=0}^{N_{\text{sim}}} \beta_t = 0.02$. 
Also, $\beta_0$ is negative as initially the suboptimality index is lowered due to the increase in the terminal cost. 

Regarding Fig.~\ref{fig:alpha-3}--\ref{fig:beta-summed-3} it can be seen that the overall expected performance is at least $\bar{\alpha}(0) +\bar{a}+ \sum_{t=0}^{\infty} \beta_t \approx 0.71 + 0.3 -0.12 = 0.89$, which lower bounds the true suboptimality. 
The deviation of the expected and true suboptimality can be explained by the computation of $\beta_t$ using $\bar{V}_N$ for the subsequent state instead of $\mathcal{V}_N$ as well as the initial jumps which come with the decrease and subsequent increase of the suboptimality index by updating the terminal cost at time step $t=0$. 


It can be observed in Fig.~\ref{fig:beta-summed-1} and \ref{fig:beta-summed-3} is that the terminal expected performance can already be predicted after two time steps once the spread in $\beta_t$ reduces. 

\begin{figure}[H]
	\centering
	\begin{subfigure}[t]{0.49\columnwidth}
		\centering
		\includegraphics[width = \columnwidth]{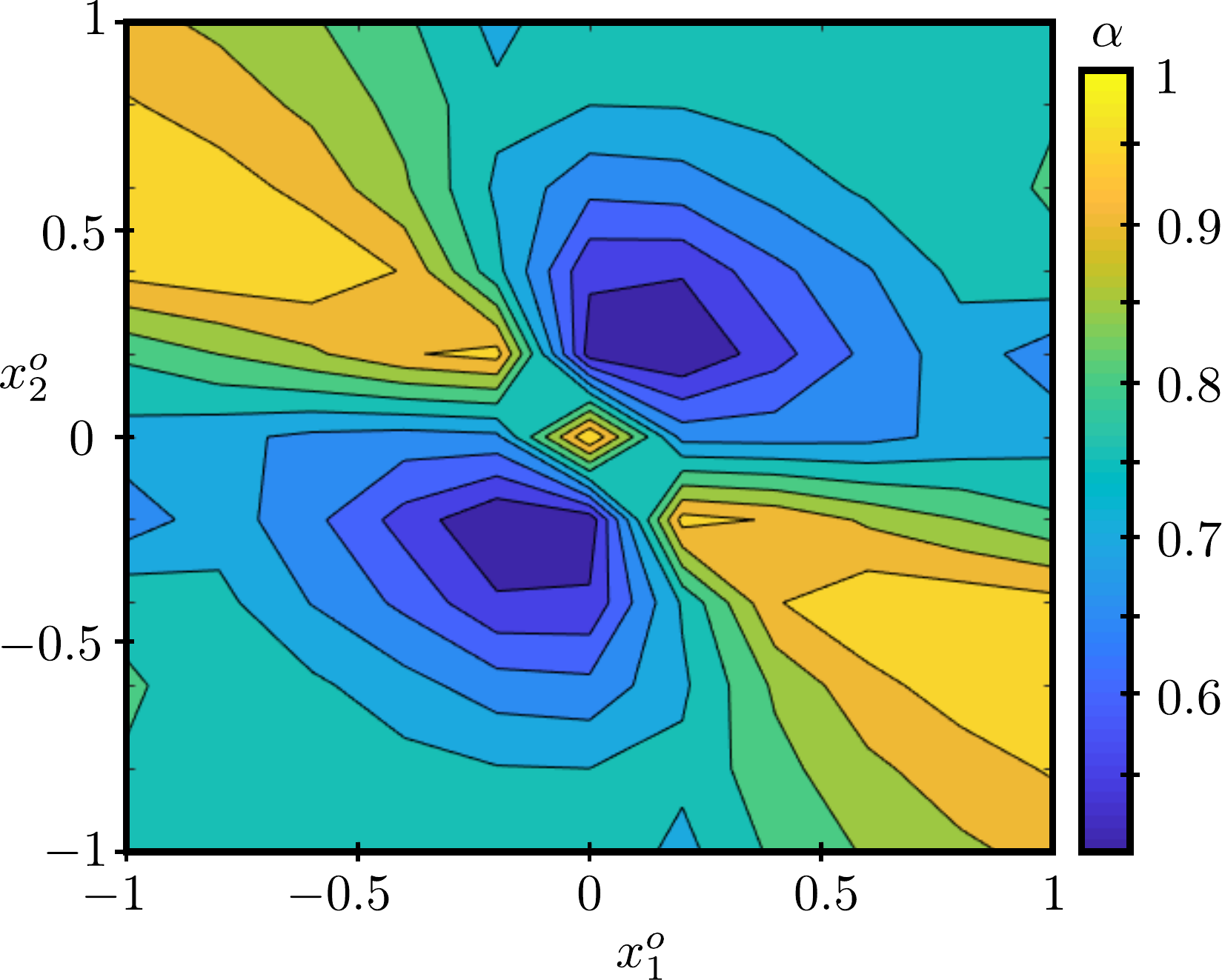}
		\caption{Cost comparison of IH optimal cost vs. MLVI-MPC.}
	\end{subfigure}
	\hfill
	\begin{subfigure}[t]{0.49\columnwidth}
		\centering
		\includegraphics[width = \columnwidth]{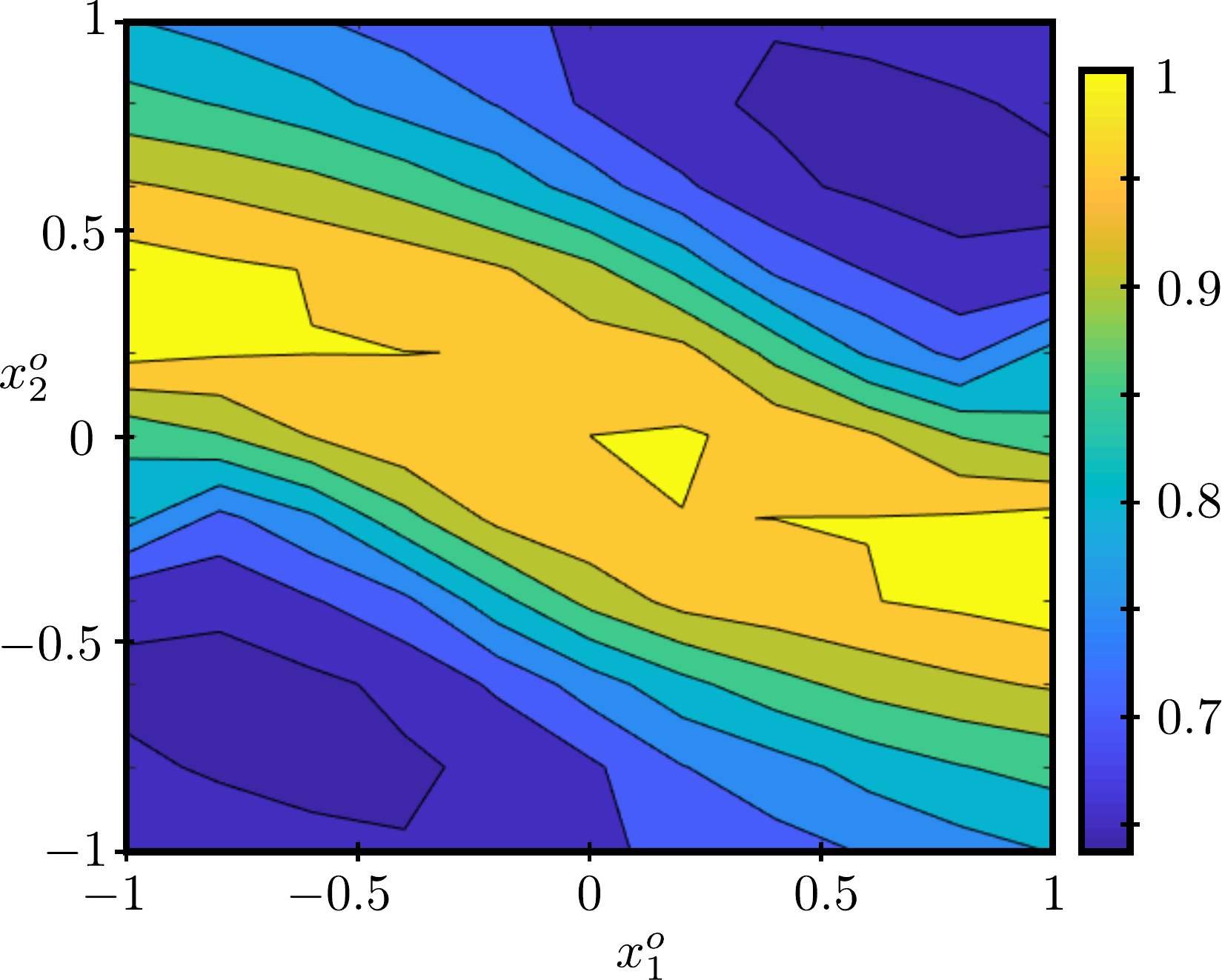}
		\caption{Cost comparison of MLVI-MPC vs. MPC.}
	\end{subfigure}
	\caption{Local cost comparison with stabilizing $N=3$.}
	\label{fig:cost-comp-N3}
\end{figure}

In order to stress the value of a terminal cost (adaptation), Fig.~\ref{fig:cost-comp-N3} shows the performance of $\kappa_N$ for a shorter horizon $N=3$ on $x^o \in [-1,1]^2$. 

While the IH cost under $\kappa_3$ is in general worse than for $\kappa_4$, as can be deduced comparing Fig.~\ref{fig:cost-comp-N3} to Fig.~\ref{fig:cost_comp_alpha}, Alg.~\ref{alg:mlvi-mpc} accomplishes vast performance improvements against MPC with short horizon. 
Although this may be somewhat intuitive, it presents the benefits of investing computational power in further online calculations. 
For example, for $N = 3$ and starting at $x^o=[-1,1]^\top$, MLVI-MPC and MPC require approx. 5.21 and 2.92 seconds, respectively, for a total of $N_{\text{sim}}$ MPC-adaptation-estimation cycles. 
The computations were performed on an IntelCore i5-10210U CPU at 1.6GHz using a straight-forward Matlab implementation without further optimization of the code or numerics.  

\section{Discussion and Conclusion}
\label{sec:conclusion}

This work investigates time-variant predictive control schemes in which the terminal costs are adapted online.
Terminal cost adaptation techniques can be found in ADP and other learning-based optimal control approaches.
The presented analysis provides bounds on the IH performance using a history stack of decay rates in convoluted form.
Moreover, improved estimates of the performance bounds can be derived online using information on the current decay rate.
It is also shown that due to stability requirements and by choice of the FH cost upper bound, the terminal cost cannot be adapted faster than the FH optimal cost decays – an issue that may be addressed in a future work.

The presented results provide a basis for assessing the influence of approximation errors in value iteration and / or terminal cost adaptation.
Furthermore, with suitable modifications the analysis may be applicable to disturbed systems, which is another future research direction. 

\bibliographystyle{plain}
\bibliography{bib/MPC,bib/ClassicOptControl,bib/ADP_RL}

\end{document}